# A multi-frequency study of the spectral index distribution in the SNR CTB 80

G. Castelletti⋆ and G. Dubner⋆⋆

Instituto de Astronomía y Física del Espacio (IAFE), CC 67, Suc. 28, 1428 Buenos Aires, Argentina
e-mail: `gcastell@iafe.uba.ar`

**Abstract.** We have conducted a study at radio wavelengths of the spectral behaviour of the supernova remnant (SNR) CTB 80. Based on an homogenised data set of integrated flux densities, we calculated for the whole SNR a radio index $\alpha = -0.36 \pm 0.02$. The shape of the global spectrum suggests absorption by ionized gas in the interstellar medium (ISM) along the line of sight. Spatial spectral variations across the SNR are investigated based on high-angular resolution data at 240, 324, 610, and 1380 MHz using different techniques. The three extended arms associated with this SNR show a clear indication of spectral steepening when moving outwards from the central nebula, with variations of up to $\Delta\alpha \sim -0.9$. However, while the spectral steepening is smooth along the eastern arm, the northern and soutwestern arms include locally flatter structures, which in all cases coincide with radio, IR and optical emission enhancements. We interpret this spectral property as the result of the combination of two different particle populations: aging relativistic electrons injected by PSR B1951+32 and particles accelerated at the sites where the SNR shock front encounters interstellar gas inhomogeneities. Concerning the central nebula, the angular resolution of the available database does not permit a detailed spectral study of the core region, i.e. the 45″ region around PSR B1951+32, where we can only confirm an average spectral index $\alpha = 0.0$. The surrounding 8′ plateau nebula has an $\langle\alpha\rangle \sim -0.25$, with a peak of $\alpha \sim -0.29$ coincident with a secondary maximun located at the termination of a twisted filament that trails to the east, behind the pulsar.

## 1. Introduction

Synchrotron radio emission is a critical tracer of particle acceleration processes in supernova remnants (SNRs), directly depending upon the energy spectrum of the relativistic electrons accelerated by the blast wave. From the study of the spatial variations in the radio spectral index $\alpha$ (where $S \propto \nu^\alpha$) it is possible to identify regions where the particles are accelerated, thus contributing to our understanding of the underlying physical processes that influence the energy of the particles. The identification of such variations permits the recognition of distinct components in a SNR (which are not always evident as distinct morphological features) and the analysis of the effects of interstellar gas inhomogeneities on the acceleration mechanisms.

CTB 80 is a Galactic SNR with a hybrid radio morphology, consisting of a flat spectrum central nebula, referred to as the plateau region ($\sim 8' \times 4'$ in size), and three extended arms, each about 30′ long, pointing to the east, north and southwest, that intersect at the center. On the western end of the central emission plateau nebula, there is a compact radio source named "the core" (size $\sim$45″) which hosts the 64 kyr pulsar PSR B1951+32 (Strom 1987; Kulkarni et al. 1988; Migliazzo et al. 2002).

Castelletti et al. (2003, hereinafter Paper I) have presented new multi-frequency radio images of CTB 80 in its full extent, based on data obtained using the Very Large Array (VLA) and the Giant Metrewave Radio Telescope (GMRT). A variety of radio features and interesting multi–frequency correlations were revealed in this study.

This paper attempts to analyze the spectral properties of CTB 80 and to establish connections between the various radio features and the local spectrum. The goal is to advance our understanding of the complex nature of CTB 80 and, in general, our comprehension of the energetic coupling between a pulsar and the host SNR.

## 2. The shape of the global spectrum

Flux density estimates of CTB 80 have been published covering a wide spectral range, from as low as 83 MHz up to 10 GHz (see Mantovani et al. 1985 and Paper I for a compilation of previous results). In Paper I we presented a preliminary plot $S(\nu)$ vs. $\nu$ to compare the newly measured flux densities at different frequencies with existing determinations. Here we perform an analysis of the global spectrum.

---

⋆ Doctoral Fellow of CONICET, Argentina.
⋆⋆ Member of the Carrera del Investigador Científico of CONICET, Argentina.

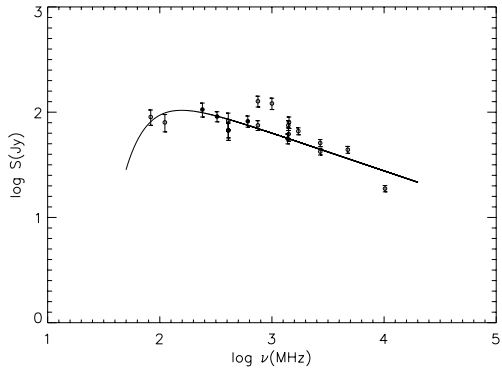

**Fig. 1.** Radio spectrum of CTB 80 from published flux densities measurements (Pauliny-Toth et al. 1966; Galt & Kennedy 1968; Velusamy & Kundu 1974; Velusamy et al. 1976; Felli et al. 1977; Haslam et al. 1982; Sofue et al. 1983; Mantovani et al. 1985; Kovalenko et al. 1994; Castelletti et al. 2003). The solid line represents the best fit to the points showing evidence for a low frequency turnover probably due to absorption by ionized gas.

In Fig. 1 we plot all published flux densities with their respective errors as quoted by the authors. All the measurements have been placed on the same absolute flux density scale given in Baars et al. (1977). They are not, however, corrected for differences in angular resolution or sensitivity of the different studies. Due to the large size and complex structure of CTB 80, a faithful estimate of the total flux density requires sensitive observations that cover the whole extent of the source, with sufficient dynamic range to reproduce structure from the very bright core to the faint emission at the borders. To take these aspects into consideration, a weighted least square fit was made, with weighting factors that take into account the respective sensitivities, angular resolutions and incomplete coverage of the source (e.g. as is the case for the 10 GHz observations). In Fig. 1 we have included the best fit to the collected data using the following spectral form

$$S_\nu = S_{240} \left(\frac{\nu}{240}\right)^\alpha \exp\left[-\tau_{240}\left(\frac{\nu}{240}\right)^{-2.1}\right] \quad (1)$$

where $S_\nu$ (Jy) represents the integrated flux density at the frequency $\nu$(MHz), and $S_{240}$ and $\tau_{240}$ are the flux density and optical depth at the reference frequency of 240 MHz. In this fitting we have not included the total flux density estimate at 178 MHz presented in Bennett (1963) because, as was noted by Mantovani et al. (1985), with the beam used for these observations the intense radio source Cygnus A is only 5.7 HPBW in RA and 1.7 HPBW in dec from CTB 80, thus contaminating the flux density estimate. The flux density estimates at 30 and 40 cm (about 1000 and 750 MHz, respectively) (Velusamy et al. 1976) lie more than $3\sigma$ from the best fit line. Few details concerning these observations are provided by the authors that could serve to understand this apparent overestimate of flux density.

The best fit to the observed flux densities is obtained with a spectral index $\alpha = -0.36 \pm 0.02$, in very good agreement with previous results ($\alpha = -0.36 \pm 0.07$, Mantovani et al. 1985).

The present fittings suggest a spectral turnover at low frequencies. As was shown by Kassim (1989), approximately two thirds of Galactic SNRs display spectral turnovers at frequencies below 100 MHz, produced by free-free thermal absorption.

From Fig. 1 it is evident that the spectral shape differs from that of the Galactic synchrotron spectrum. In effect, if the radio emission from CTB 80 were simply due to shock compression of the ambient Galactic magnetic field and the ambient population of relativistic electrons, then the radio spectrum of the SNR should resemble the Galactic spectrum (i.e. $\alpha \sim -0.5$ and $\sim -0.8$ below and above 400 MHz, respectively; Salter & Brown 1988). The observed spectral behavior suggests that other factors control the radio spectrum in CTB 80.

From Eq. (1) we have also derived the best-fit value of the free-free optical depth at 240 MHz, $\tau_{240} = 0.07$ (which corresponds to $\tau_{74} \sim 0.8$ when scaled to 74 MHz assuming a $\nu^{-2.1}$ dependence). For a source distant about 2 kpc this optical depth would imply the presence of a thermal plasma (assumed homogeneous) with an electron density $N_e \sim 2$ cm$^{-3}$. This value for $N_e$ is obtained from the relation $\tau = 8.235 \times 10^{-2} \nu^{-2.1} T_e^{-1.35} EM\, a(\nu, T)$ (Mezger & Henderson 1967), where $T_e(K)$ is the electron temperature in Kelvin (assumed to be about 8000 K, for a typical warm ionized medium (WIM), Haffner et al. 1999), $\nu$ is the frequency in GHz, EM is the emission measure ($EM = \int N_e^2\, dl$ cm$^{-6}$ pc), and $a(\nu, T)$ is the Gaunt factor $\sim 1$ (Rohlfs 1990). However, this derived value for $N_e$ is at least an order of magnitude higher than typical WIM electron densities (Reynolds 2004).

As discussed by Kassim (1989), the estimates of the optical depth can be highly uncertain if the low-frequency turnover occurs near the lowest observing frequency, resulting in its overestimation. In a study of the SNR W49B, Lacey et al. (2001) have found high optical depth, varying across the source from $\tau_{74} = 0.7$ to the east to $\tau_{74} = 1.6$ to the west of the SNR. The authors discuss intrinsic and extrinsic absorption processes and conclude that low density gas from an "extended HII region envelope (EHE)" is responsible for the low frequency absorption. We can speculate as to whether a similar model could apply to the environment of the SNR CTB 80. An intervening medium with an electron density $N_e \sim 2$ cm$^{-3}$, as estimated above, would imply a dispersion measure for the pulsar PSR B1951+32 of $DM \sim 4000$ cm$^{-3}$ pc, about two orders of magnitude higher than the observed value, $DM = 45$ cm$^{-3}$ pc (Kulkarni et al. 1988). A possible way to account for this discrepancy would be to assume that the absorbing plasma is not uniformly distributed, but confined to a narrow ionized sheet (barely $\sim 0.2$ pc in thickness) located between us and CTB 80. Such a sheet would need to cover a region of a few degrees in the sky (the nearby source Cygnus A does not show the same effect). Accurate flux density measurements of CTB 80 at very low frequencies ($\leq 80$ MHz) are highly desirable to establish the frequency dependence of the spectrum in this region and to understand the properties of the ionized gas in this direction of the Galaxy.

## 3. Spatial variations of the spectral index across CTB 80

The study of spatial variations of the spectrum across CTB 80 is a sensitive tool for understanding the coupling between the fresh relativistic electrons constantly supplied by the pulsar and the shocked plasma in the three long arms of this SNR. It serves also to probe whether the three extended arms of this abnormally shaped SNR belong to a unique source (in the case that all of them show similar spectral characteristics) or whether they are the consequence of the superposition of different emissions along the line of sight.

To carry out this study over the extended emission of CTB 80 we have used good quality observations at three frequencies: 240, 610, and 1380 MHz. The observations at 240 and at 1380 MHz were obtained with the GMRT and VLA synthesis telescopes, respectively. Observational details are presented in Paper I. The data at 610 MHz were obtained with WSRT (HPBW $50'' \times 89''$, Strom private communication). Special care was taken to correct the data for missing spatial frequencies in order to accurately determine the remnant's flux density at all frequencies. Also, because the data at the three frequencies considered come from three different radio telescopes, all data sets were carefully matched before the combination. To avoid a positional offset, all images were aligned to the same pixel position and convolved to the same spatial resolution. The final common beam is $50'' \times 90''$, PA = 77° for the 240–610 MHz comparison, and $93'' \times 78''$, PA = 72° for the 610 and 1380 MHz bands. Also, differences in flux density scales between the different sets of observations were considered and corrected.

Figure 2 shows the spectral index distribution obtained between 610 and 1380 MHz, with a few contours of the total flux density at 1380 MHz superimposed for reference. The map contains two spectral artifacts near ($19^h52^m16^s$, $32°50'28''$) and around ($19^h54^m6^s$, $33°17'42''$), produced by the presence of unrelated bright point sources. These will not be considered in the following analysis. In this greyscale representation of the spectral index, darker means flatter spectrum. Two aspects can be especially noted from this figure: (a) the bright central nebula, core and plateau regions show, as expected, the flattest spectrum, with a mean spectral index $\langle\alpha\rangle = -0.25\pm0.05$, while along the low-brightness extended components of the remnant, a clear steepening of the spectrum with increasing distances from the pulsar is observed (see also Fig. 3); this behavior is reminiscent of that seen in extragalactic radio jets associated with active radio galaxies (Katz-Stone & Rudnick 1997), and (b) there is good general correspondence between morphological features and the spatial distribution of the radio spectral index in the sense that brighter radio features tend to have flatter indices than their surroundings. Note for example the extension protruding to the north from the central nebula, near ~$19^h53^m$, $33°00'$ (called the "northern protrusion", NP, in Paper I) and the coincidences in the interior of the northern arm. Also striking is the flattening of the spectrum along the western border of the remnant, particularly in the arc of emission westwards of the plateau nebula.

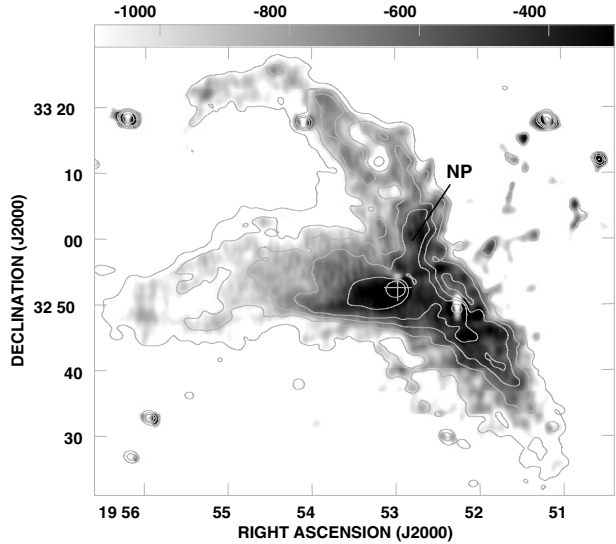

**Fig. 2.** Greyscale image of the spectral index distribution across the SNR CTB 80 as derived between 610 and 1380 MHz. The wedge displays the gray scale in units of milli − spectral index. The cross indicates the location of the pulsar PSR B1951+32. The 1380 MHz total flux contours plotted for reference are traced at 30, 50, 70, 95, 200 and 400 mJy beam$^{-1}$.

The possibility that the north and southwest arms resulted from the injection of pulsar-generated relativistic particles into an old shell's compressed magnetic field was originally proposed by Fesen et al. (1988) based on the analysis of infrared features. In general, the present spectral index map, traced with very good angular resolution and sensitivity, is consistent with this picture. However, not all the flatter spectrum regions observed around the pulsar (towards the north and southwest arms of CTB 80) are spectrally connected as a continuum with the central nebula. Narrow bands of steeper index ($\alpha \sim -0.5$) divide the central plateau from the NP, the westward arc ahead of the pulsar and the bright features in the southern arm. The radiocontinuum enhancements accompanied by spectral flattening observed in the north and southwest arms appear more likely related to local conditions, such as the presence of radiative shocks, or, a mixture of the pulsar's influence and local characteristics of the host SNR.

On the basis of infrared data, Fesen et al. (1988) also proposed that the eastern arm is produced by the pulsar's relativistic particle wake inside a cavity created by the SNR. In this case, a smooth spectral steepening is observed, consistent with this hypothesis.

To analyze in detail the central-nebula/arms spectral transitions, in Fig. 3 we plot the spectral index as a function of the distance from the core along the three arms of the remnant. The plotted points correspond to the spectral indices averaged over annuli of width 10″, within three angular sectors that approximately cover the three arms of CTB 80. The spectrum gradually steepens along each branch beyond the central component, with extreme values of $\alpha \sim -0.96$. Such steepening, however, presents different characteristics along each arm. The fastest steepening occurs along the northern arm, where $\alpha \sim -0.8$ at a distance of about 600″ away from the pulsar, in contrast with $\alpha \sim -0.5$ measured on the eastern and southwestern components at the same distance. The eastern arm has two peaks of flat spectrum corresponding to structures localized within the plateau nebula. However, away from the central nebula the variation of $\alpha$ is smooth. Our results are compatible with Sofue et al.'s (1983) analysis based on the comparison between 2.7 and 10 GHz data. On the northern arm, the two peaks with spectral indices flatter than their surroundings are placed at about 900″ and 1200″ from the core nebula, and correspond to the location of the peculiar branching observed in this arm. These regions with a flatter spectrum possibly indicate higher compression experienced by the emitting plasma due to interaction of the blast wave with denser ambient medium.

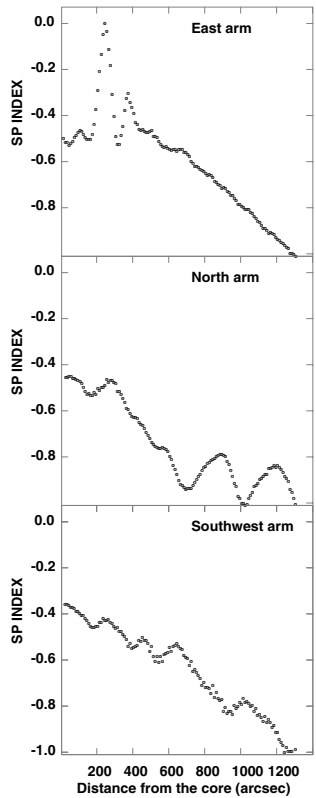

**Fig. 3.** Spectral indices calculated between 610 and 1380 MHz as a function of distance from the central nebula within angular sectors that include the eastern (*top*), northern (*middle*), and southwestern (*bottom*) arms of the SNR CTB 80.

### 3.1. Spectral distribution from analysis boxes

The spatial variation of $\alpha$ across the remnant was also investigated through the combination of 240, 610, and 1380 MHz data in two frequency pairs: 240−610 and 610−1380 MHz, in order to investigate the curvature of the spectrum.

To carry out this study we used T−T plots (Costain 1960; Turtle et al. 1962). This method is based on the plot of brightness temperature at one frequency against the corresponding one at the other frequency, within a chosen spatial region. The slope of the linear function that produces the best fit to the data is related to the spectral index between these two frequencies. This procedure permits accurate spectral index determinations even if differences in the zero-levels between the images are present. Background variations smaller than the size of the selected region are taken into account in the uncertainty of the linear regression.

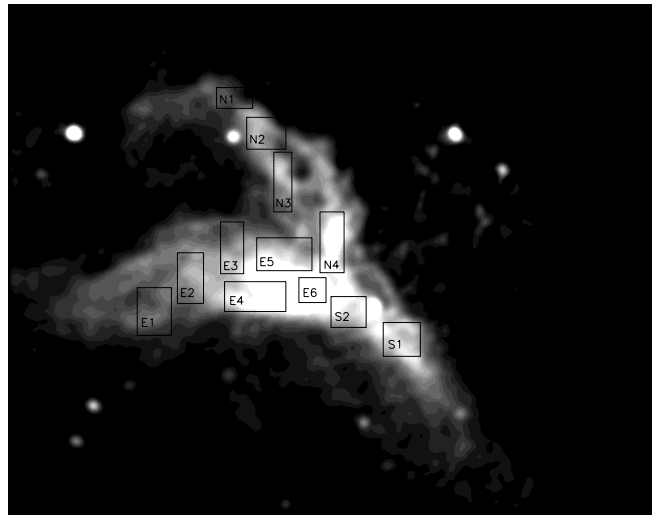

**Fig. 4.** The boxes used for spectral index calculations using T−T plots displayed on a gray scale image of the extended emission of SNR CTB 80 at 1380 MHz.

We have applied the T−T plot method in twelve different regions of the remnant, selected to cover the main features of this peculiar SNR. The regions were chosen with the criteria that: (a) they include the central plateau and the three outer arms of CTB 80; (b) they do not include background point sources and regions with poor signal-to-noise ratio (close to the borders) at any of the considered frequencies; and (c) they include sufficient pixels to provide good statistics. In Fig. 4, overlapping the total intensity image of CTB 80 at 1380 MHz, we show the selected regions where the technique was applied. The labels N, E, and S in the areas identification refer to the north, east or southwest arms in CTB 80, respectively. As can be seen from Fig. 4, the analysis boxes have different sizes, with areas that vary from $\sim 5 \times 10^4$ to $\sim 18 \times 10^4$ arcsec$^2$ (which means between 11 and 40 beams for the 240−610 MHz comparison, and between 7 and 25 beams for the 610−1380 MHz frequency pair).

To get a better estimate of the spectral index, the regression was done twice, by assuming X vs. Y and vice-versa. We obtain the final spectral index measure from the average slope of both, with uncertainties that have a statistical origin from the straight line fits plus the above-mentioned flux density scale corrections. The pulsar wind nebula is separately analyzed in the next section. Figure 5 displays a few sample T−T plots, and the resulting average fitted line. To facilitate the comparison between different regions and frequency pairs, in Fig. 5 we have used the same scales. Table 1 summarizes the radio spectral indices obtained for each region, together with their errors derived using this method.

Several important conclusions can be drawn from this analysis. The results summarized in Table 1 confirm the spectral steepening in all directions away from the central nebula as is qualitatively shown in Figs. 2 and 3. Regions E4, E5, and E6, over the central plateau, have, as expected, flat spectra over the whole range of frequencies; the flatter the spectrum, the closer the region is to the pulsar. As noted above, the region N4, coincident with the "northern protrusion", has an average index steeper than the plateau, but flatter than its surroundings. This fact is consistent with the presence of radiative optical filaments and intense infrared radiation associated with this feature (Fesen et al. 1988; Mavromatakis et al. 2001). A likely explanation is that at this spot the expanding blast wave has encountered a denser interstellar clump and locally generated radiative shocks with stronger post compression shock densities and magnetic field strength (Katz-Stone et al. 2000). The observed $\alpha$ implies a strong density jump, with a shock compression factor $r \sim 5.2$, as obtained based on the equation $\alpha = \frac{3}{2(r-1)}$, under the linear diffussive shock acceleration theory (Reynolds 1988).

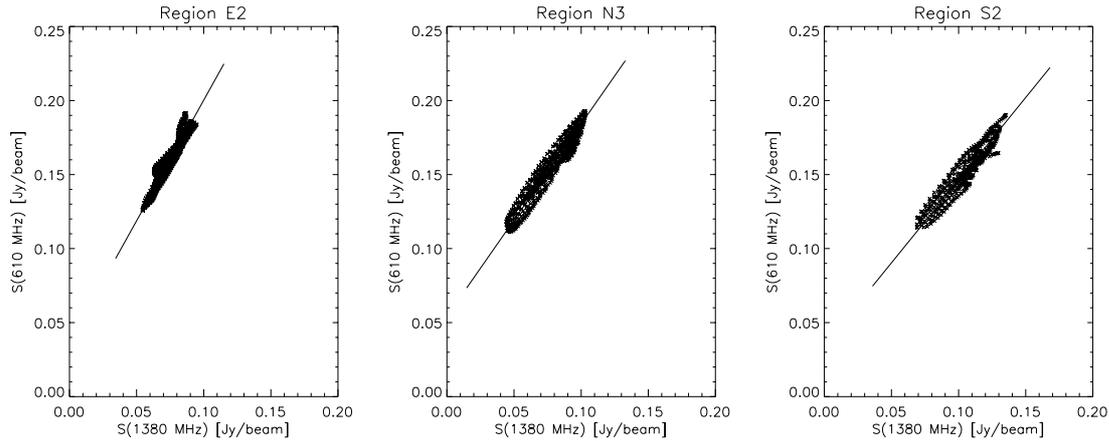

**Fig. 5.** Sample T–T plots between 610 and 1380 MHz. The caption above each frame corresponds to the identifications shown in Fig. 4. The solid lines represent the least-square fit to the points.

**Table 1.** Radio spectral index distribution over CTB 80.

| | Box labels | | | | | |
|---|---|---|---|---|---|---|
| Eastern branch | E1 | E2 | E3 | E4 | E5 | E6 |
| $\alpha_{240-610}$ | $-0.72 \pm 0.12$ | $-0.61 \pm 0.12$ | $-0.28 \pm 0.09$ | $-0.09 \pm 0.08$ | $-0.35 \pm 0.09$ | $-0.03 \pm 0.08$ |
| $\alpha_{610-1380}$ | $-0.72 \pm 0.04$ | $-0.60 \pm 0.03$ | $-0.23 \pm 0.03$ | $-0.11 \pm 0.03$ | $-0.43 \pm 0.03$ | $-0.11 \pm 0.04$ |
| Northern branch | N1 | N2 | N3 | N4 | | |
| $\alpha_{240-610}$ | $-0.75 \pm 0.16$ | $-0.41 \pm 0.09$ | $-0.96 \pm 0.08$ | $-0.48 \pm 0.08$ | | |
| $\alpha_{610-1380}$ | $-0.10 \pm 0.08$ | $-0.17 \pm 0.03$ | $-0.31 \pm 0.04$ | $-0.36 \pm 0.04$ | | |
| Southern branch | S1 | S2 | | | | |
| $\alpha_{240-610}$ | $-0.81 \pm 0.08$ | $-0.72 \pm 0.11$ | | | | |
| $\alpha_{610-1380}$ | $-0.56 \pm 0.06$ | $-0.22 \pm 0.05$ | | | | |

Along the eastern arm, while the spectral indices have large spatial variations for a given frequency pair, the continuum spectrum at any point appears to be essentially straight between 240 and 1380 MHz. In contrast, the northern and southwestern extensions show notable changes in $\alpha$ between the frequency pairs at any given position. All these regions are found to have concave radio spectra, flattening towards higher frequencies.

As pointed out by Leahy & Roger (1998) a concave spectrum can be produced by a similarly curved electron energy spectrum, or by two different electron populations. In the Cygnus Loop SNR, it has been found that concave spectra are found in regions dominated by diffuse emission, while convex spectra are found to be associated with sharp radio filaments on the borders of the SNR (Leahy & Roger 1998). This kind of limb brightened emission is absent in CTB 80 where generally the boundaries fade smooth away. We propose that the synchrotron emission originating in the N and SW arms has two components: electrons freshly injected by the pulsar, and particles accelerated at the shock front.

## 4. The central nebula

The central nebula of CTB 80 was described by Angerhofer et al. (1981) as formed by two spectrally different components: the "plateau", an elongated feature about 10′ long, with a mean spectral index $\alpha = -0.33$ and the "core", a circular compact component located at the western edge of the plateau, about 0.′5 in size, with $\alpha = 0.0$. The presence of a pulsar in the core was later proposed and confirmed (Strom 1987; Kulkarni et al. 1988).

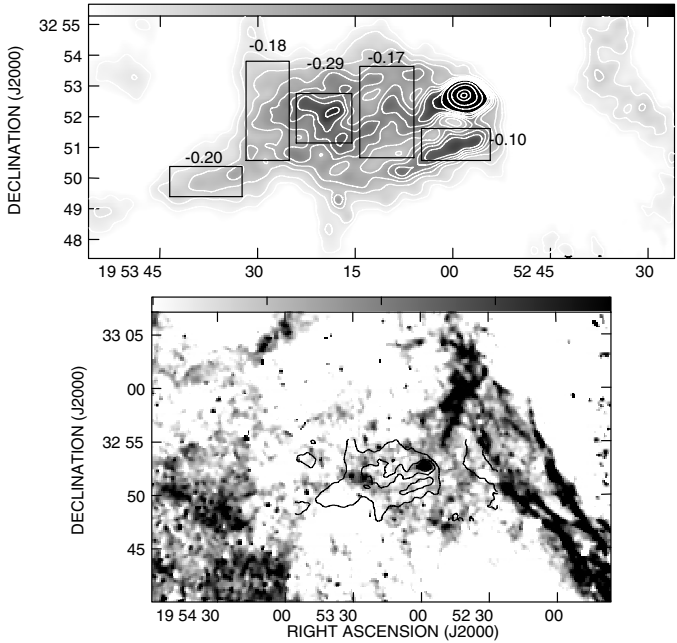

**Fig. 6.** *Top*: the plateau nebula at 618 MHz (Paper I) with the boxes used for spectral index estimates. The numbers correspond to the local spectral indices as obtained from the comparison between images at 324 and 1380 MHz. *Bottom*: optical image in the $H_\alpha$ line of the central region of CTB 80 (From Mavromatakis et al. 2001), with a few radio contours showing the plateau nebula at 618 MHz.

Angerhofer et al. (1981) described the plateau as a nebula having nearly constant surface brightness, but containing the above–mentioned compact core and a secondary emission peak near the eastern border. New observations of the central nebula carried out at 618 MHz using the GMRT (about 8″ HPBW, Paper I) revealed that the bright plateau nebula in CTB 80 is not smooth but structured. The most prominent feature is a twisted filament, about 6′ along, which runs in an E−W direction, starting at the bright compact core formed around the pulsar. The secondary peak reported by Angerhofer et al. (1981) is located at the termination of the twisted feature (See for example Fig. 6, top). At the angular resolution of our multi-spectral database, the core is not resolved, and based on our data we can only confirm its very flat spectrum.

To investigate the spectrum of the plateau we made a comparison between the VLA images at 324 and 1380 MHz. It is useful to make the analysis based on an homogeneous database and over a relatively broad spectral range. We did not use the high-resolution, GMRT 618 MHz image in this calculation because, as discussed in Paper I, short spatial frequencies are missing and the flux recovery is imperfect. The images at 324 and at 1380 MHz were reduced to the same geometry and angular resolution. Over this frequency range, the average spectral index of the plateau is $\alpha_{324}^{1380} = -0.28 \pm 0.02$, consistent within the errors with the value $\alpha_{610}^{1380} = -0.25 \pm 0.05$, previously derived based on a comparison using the WSRT 610 MHz image (Sect. 3).

To carry out a detailed analysis, we have considered different regions containing features of interest and calculated the spectral index using direct comparison between the images, and T−T plots. The selected boxes are shown in Fig. 6 (top) superimposed on the high-resolution 618 MHz image of the plateau nebula. From this study we derived the following average spectral indices: $-0.10 \pm 0.05$, $-0.17 \pm 0.10$, $-0.29 \pm 0.05$, $-0.18 \pm 0.06$, $-0.20 \pm 0.06$ from west to east, confirming the existence of a steepening in the spectrum when moving away from the pulsar region to the tail of the plateau nebula. However, the steepest index is not attained at the eastern edge of the nebula, but coinciding with the secondary maximum, near $19^h53^m19^s$, $32°52'$. Figure 6 (bottom) shows a comparison of the plateau nebula as observed at 618 MHz (black contours) with the $H_\alpha$ image of the central region of CTB 80 (Mavromatakis et al. 2001). It is suggestive that a knot of $H_\alpha$ emission from recombining hydrogen lies exactly at the termination of the twisted filament, where the secondary maximum forms. Although it is not possible to separate emission originating within the plateau from background $H_\alpha$ contribution, based on the observed morphological agreements it can be speculated that the $H_\alpha$ knot represents a shocked interstellar cloudlet. If this is the case, then the steeper spectrum observed at the secondary maximum is the result of a mixture of electron populations, i.e. the aging relativistic electrons injected by the pulsar and shock accelerated particles at the site where the pulsar "tail" encounters an interstellar clump.

## 5. Discussion and conclusions

We have recalculated the global spectral index of the SNR CTB 80 on the basis of an homogenised multi–frequency data base, where the differing observational parameters of all the published data were taken into consideration. A weighted fit produces a spectral index $\alpha = -0.36 \pm 0.02$ for the whole SNR. At frequencies below ∼200 MHz, the effects of absorption by a thermal plasma are observed.

We have also carried out a detailed study of the radio spectral index variations across CTB 80 using different techniques. From this study we can conclude that the radio spectra of the three extended components of CTB 80 steepen when moving away from the pulsar position, with variations of up to $\Delta\alpha \sim -0.9$. The behaviour, however, is not identical, the eastern arm exhibits a smooth, regular steepening, whereas the north and southwestern arms include regions with local spectral flattening which in all cases coincide with radio, infrared and optical emission enhancements. The continuity of the spectral index distribution confirms that CTB 80's components belong to a single object.

We have also analyzed the curvature of the spectrum based on data at three different radio frequencies. Our results reveal that along the eastern component there is almost no variations with frequency. In contrast, along the north and southwest arms, all the investigated regions have a concave spectrum.

On the basis of the present results, we can conclude that to the east the dominant source of relativistic electrons must be the pulsar PSR B1951+32, and the observed steepening must be due to the energy dependence of the decay time of the relativistic electrons, as earlier proposed by Hester & Kulkarni (1988). For the north and southwest arms, different processes seem to be taking place: injection of relativistic electrons by the pulsar and particle acceleration at the shock front, particularly at the sites where the expanding shock encounters interstellar inhomogeneities. If these results are interpreted within the frame of simple diffusive shock wave acceleration processes in the test particle limit, then the observed spectral variations are associated with different compression ratios in the SNR's arms. The steep spectrum observed in the extremes of the arms can be interpreted as evidence of weaker and older shocks (lower Mach numbers).

The analysis of the central plateau revealed that the spectral index is not uniform, but smoothly steepens from $\alpha = 0$ in the core to $\alpha \sim -0.2$ at the eastern end. Within the nebula the feature with the steepest spectrum ($\alpha \sim -0.29$) is the secondary maximum located at the termination of the twisted filament that trails behind the pulsar core. The proximity and morphological concordance observed between this radio maximum and an $H_\alpha$-bright knot suggests an interaction with an interstellar clump at this site. Therefore, the derived spectral index might be reflecting the presence of two different particle populations, the freshly accelerated electrons injected by PSR B1951+32 and particles accelerated at the site where radiative cooling occurs.

*Acknowledgements.* We are grateful to Richard Strom for providing us with the 610 MHz WSRT image and to Fotis Mavromatakis for the optical images. We wish to thank the referee, Dr. Chris Salter, whose constructive criticism has helped to improve this paper. This work was supported by the grants CONICET 2136/00, ANPCyT-PICT 03-14018 and UBACYT A055/04.